# Transmissive Metagrating for Arbitrary Wavefront Shaping Over the Full Visible Spectrum

Zi-Lan Deng[1,*,†], Xuan Ye[1,†], Hao-Yang Qiu[2,†], Qing-An Tu[1], Tan Shi[1], Ze-Peng Zhuang[2], Yaoyu Cao[1], Bai-Ou Guan[1], Naixing Feng[3], Guo Ping Wang[3], Andrea Alù[4], Jian-Wen Dong[2,*], Xiangping Li[1,*]

[1]Guangdong Provincial Key Laboratory of Optical Fiber Sensing and Communications, Institute of Photonics Technology, Jinan University, Guangzhou 510632, China.
[2]School of Physics and State Key Laboratory of Optoelectronic Materials and Technologies, Sun Yat-sen University, Guangzhou 510275, China.
[3]Institute of Microscale Optoelectronics, Shenzhen University, Shenzhen 518060, China.
[4]Photonics Initiative, Advanced Science Research Center, City University of New York, New York, NY 10031, USA

## Abstract

Metagratings have been shown to form an agile and efficient platform for extreme wavefront manipulation, going beyond the limitations of gradient metasurfaces. Previous approaches for transmissive metagratings have resorted on compound asymmetric inclusions to achieve single-channel near-perfect diffraction. However, such complex inclusions are sensitive to geometric parameters and lack the flexibility for arbitrary phase modulation, restricting applications to beam deflection. Here, we show perfect unitary diffraction in all-dielectric transmissive metagratings using rectangular inclusions by tailoring their multipole interferences. Using this principle, we experimentally demonstrate analog phase profile encoding of a hologram through displacement modulation of CMOS-compatible silicon nitride nanobars, manifesting broadband and wide-angle high diffraction efficiencies for both polarizations and across the entire visible range. Featured with extreme angle/wavelength/polarization tolerance and alleviated structural complexity for both design and fabrication, our demonstration unlocks the full potential of metagrating-based wavefront manipulation for a variety of practical applications.

*E-mail: zilandeng@jnu.edu.cn, dongjwen@mail.sysu.edu.cn, xiangpingli@jnu.edu.cn

†These authors contributed equally to this work.

All-dielectric metasurfaces with discrete Mie scattering inclusions are recognized as an efficient platform to manipulate the light wavefront due to their negligible losses compared with plasmonic counterparts[1]. Geometric Pancharatnam-Berry phase[2], propagating phase[3, 4] and their hybridization[5, 6] have been widely employed in gradient metasurfaces to efficiently control the phase, polarization and amplitude profiles of light wavefront, leading to multifunctional planar optical devices such as achromatic metalenses[7-9], high-quality meta-holograms[10-12], tunable meta-devices[13] and mixed hologram and color printing[14, 15]. As there is a tradeoff between efficiency and spatial frequency bandwidth in gradient metasurfaces[16, 17], metagratings have emerged as an alternative platform to perform wavefront steering functionalities, especially for some extreme wavefront transformation scenarios.

In a metagrating, the scattering properties of identical inclusions in a periodic lattice are carefully tailored to achieve the desired diffraction order management, different from phase-gradient metasurfaces that employ a discrete set of different inclusions to map digitized phase levels in the 2π range. Guiding light towards a specified diffraction channel while suppressing all other diffraction orders has been the most popular application of metagratings so far. Metallic metagratings can easily achieve such high-efficiency single channel diffraction by tuning the plasmonic resonance in the extraordinary optical diffraction regime[17, 18], yet these approaches work better in reflection modes and their absolute efficiencies are hindered by intrinsic Ohmic losses[19-29]. Achieving single channel diffraction in all-dielectric transmissive metagratings is more challenging, as more diffraction channels coexist in both reflection and transmission semi-spaces. Usually, compound inclusions consisting of multiple dielectric scatters with specific asymmetric geometrics have been employed to achieve the near-perfect diffraction at the transmission side[30-39]. However, more advanced wavefront shaping functionalities typically lead to significant complexities in the design and suffer from stringent demands in fabrication complexity.

Here, we show that it is possible to achieve perfect diffraction in all-dielectric transmissive metagratings composed of simply binary inclusions, enabling highly efficient phase profile modulation over the full visible spectrum. The proposed perfect diffraction transmissive metagrating is composed of simple rectangular inclusions with neither high refractive index nor high aspect ratio. Perfect steering of light into a single diffraction order at the transmission side with suppressed specular and other higher order channel reflection or transmission is achieved through multipolar interference manipulation in all-dielectric nanobars. In addition, anomalous light channeling sustains robust efficiency within a broadband, wide-angle range and applicable for both transverse electric (TE) and transverse magnetic (TM) polarizations. Thanks to the structural simplicity of such inclusions, an arbitrary phase profile can be continuously modulated through a displacement-varying detour phase scheme.

We have experimentally realized these high-efficiency metagrating holograms with robust performance. Compared with the widely explored gradient metasurfaces and traditional binary optics, our proposed metagratings open new routes towards high-efficiency light wavefront shaping, especially in scenarios requiring analog modulation and extreme angle/wavelength/polarization tolerance. Such alleviated structural complexity makes mass production for daily life usages feasible based on low-cost fabrication techniques such as nano-imprinting.

Figure 1a shows the schematic of the proposed diffraction metagrating, made of silicon nitride ($SiN_x$) nanobars with width $w$, height $h$, periodicity $p$ and refractive index $n_g$ on top of a silicon dioxide ($SiO_2$) substrate with refractive index $n_s$. By inspecting the diffraction order chart in Fig. 1b, we can locate a pair of diffractive regimes denoted by the dotted shadow areas, where there are only $0^{th}$ and $-1^{st}$ diffraction channels in both transmission ($T_0$, $T_{-1}$) and reflection sides ($R_0$, $R_{-1}$). They are delimited by transmission Wood anomaly (TWA, blue) lines and reflection Wood anomaly (RWA, red) lines. The left target regime is delimited by $0^{th}$ TWA, $2^{nd}$ RWA, $-1^{st}$ RWA and $1^{st}$ TWA, and the right target regime is delimited by $0^{th}$ TWA, $-2^{nd}$ RWA, $1^{st}$ RWA and $-1^{st}$ TWA, respectively. The $\pm i^{th}$ (i=0, 1,2, …) TWA and RWA lines are determined by

$$k_x \pm i2\pi / p = \pm k_0, \tag{1}$$

$$k_x \pm i2\pi / p = \pm n_s k_0, \tag{2}$$

respectively, where $k_0=\omega/c$ is the wavevector of incident light in vacuum, $k_x$ is the $x$-component wavevector. In such a few-diffraction-order grating configuration, proper engineering of the interference of the multipole modes supported by the $SiN_x$ nanobars enables full transmission of the incident light towards a diffraction order of choice, including negative refraction towards the $T_{-1}$ diffraction direction, as in Fig. 1a. In order to then realize arbitrary wavefront shaping at this diffraction order, we can further modulate the metagrating using an interferometric fringe profile, as shown in Fig. 1c. This is achieved by exploiting the interference phase difference between adjacent nanobars originating from the light path difference between the incident side $d_i(x,y)=d(x,y)\sin\theta_0$ and the one in the transmission side $d_t(x,y)=d(x,y)\sin\theta_{-1}$ (Fig. 1d). Combined with the grating equation $k_0(\sin\theta_0+\sin\theta_{-1})=2\pi/p$, the modulated phase profile is then given by

$$\varphi(x, y)=k_0(d_i+d_t)=2\pi d(x, y)/p, \tag{3}$$

where $d$ is the shifted displacement of the $SiN_x$ nanobar from its original periodic position, and $p$ is the periodicity of the metagrating.

Figure 2 shows the theoretical diffraction properties of a perfect diffraction metagrating with parameters $w$=162 nm, $h$=267.5 nm, $p$=500 nm, $n_g$=2.5, $n_s$=1, obtained by rigorous coupled wave analysis (RCWA). For a TM-polarized plane wave under oblique ($\theta_0$=30°) illumination (magnetic field $\boldsymbol{H}$ pointing along the $y$-direction), we find unitary perfect diffraction at λ=532 nm, for which the diffraction efficiency in the transmissive -1$^{st}$ order reaches $T_{-1}$=100%, while all other diffraction channels ($R_0$, $R_{-1}$, $T_0$) are suppressed (Fig. 2a). The field pattern ($H_y$) shows that the oblique incident light completely goes through the metagrating and is bent in the negative refraction direction (Fig. 2b). The scattering cross-section spectra of the different multipoles supported by the $SiN_x$ nanobar (Fig. 2c) shows that the electric dipole and magnetic dipole contribution are dominant in this wavelength range, the far-field interference of these modes causes single-channel perfect diffraction. Previously reported silicon cylinder gratings have also shown to support negative refraction with high efficiency[40-42]. Although the circular cross-section allows an easy analytical treatment of the multipole expansion, it lacks the flexibility to independently tune the multipoles, leading to incomplete suppression of other diffraction orders[42]. For TE-polarized illumination (***E*** field along the *y* direction), the anomalous diffraction efficiency is also broadband, although the peak efficiency is limited to 90% (Fig. 2d). The field pattern (Fig. 2e) shows some distortion in both incident and bending wavefront, due to the interference between the main beams and the leakage into other diffraction orders. The broadband high diffraction efficiency can be attributed to the interference of the magnetic dipole and magnetic quadrupole resonance at shorter wavelengths and interference of magnetic dipole and electric dipole at longer wavelengths (Fig. 2f). The metagrating manifests high diffraction efficiency for large range of wavelengths and incident angles, as shown in Supplementary Fig. S1. Furthermore, the metagrating performance is immune to geometric imperfections (Supplementary Fig. S2), which relaxes the fabrication demand and facilitates the practical implementation significantly.

The inclusion simplicity of the binary metagrating allows for arbitrary wavefront shaping beyond beam deflection. As an example, we employ the perfect diffraction metagrating to encode the phase profile of a hologram (Supplementary Fig. S3). The phase profile (Supplementary Fig. S3a) of a horse image was calculated by the Gerchberg–Saxton (GS) algorithm[43]. Based on the displacement encoding approach denoted in Fig. 1c, we can map the phase profile continuously onto a curved grating profile (Supplementary Fig. S3b), which is similar to an interferometric fringe between the object wave and a reference plane wave. Resorting to an angle-resolved optical setup (Fig. 3a), we can capture holographic images by varying both wavelength and incident angles of the modulated metagrating. The $SiN_x$ metagrating was fabricated by electron beam lithography (EBL) following the process described in Supplementary Fig. S4. Figures 3(b-e) show scanning electron microscope (SEM) images of the metagrating sample in both top and oblique views, manifesting continuously curved $SiN_x$ segments with perfect steep flank profile, well matching the theoretically designed shapes.

Figure 4 shows experimentally reconstructed holographic images over near-full incident angle range (10°~80°) and full visible frequency range (450 nm~700 nm) for TE polarization. The holographic image sizes are different for different colors, according to the scaling factor $\lambda/z_d$ ($\lambda$ is the wavelength, $z_d$ is the distance between image and hologram plane). For clarity, we just show the minimum picture ranges containing the horse image, abandoning their relative size information. It is clear that all the reconstructed images have high fidelity over the full visible band and for a broad angular range, except those angles beyond the cut-off angle of the metagrating regime (right-lower corner of Fig. 4), which corresponds to the $-1^{th}$ TWA lines in Fig. 1b. The holographic images span over all colors, including purple, blue, cyan, green, yellow, orange and red. Although in the boundary images are a little blurred and stretched due to the reduced diffraction efficiency and enlarged diffraction beam width, most angles and wavelengths report high-fidelity holograms. In addition, the holographic images can be reconstructed with high-fidelity for both TE and TM polarizations, as shown in Supplementary Fig. S6, manifesting that the metagrating has well-performed polarization tolerance.

The diffraction efficiency of the experimentally realized metagratings are analyzed in Fig. 5. Figs. 5a and b show the calculated diffraction efficiencies ($T_{-1}$) with realistic material parameters of $SiN_x$, as shown in Supplementary Fig. S7. The other optimized parameters of the realistic metagrating are $w$=150 nm, $h$=300 nm, $p$=400 nm, and $n_s$=1.5. Figs. 5c to f show calculated and measured efficiencies of all existing diffraction orders, respectively, at a fixed incident angle $\theta_0$=60°. For TE polarization (Figs. 5c, e), the calculated diffraction efficiency of $T_{-1}$ shows a broad bandwidth of operation, from 470 nm to 730 nm, spanning the entire visible range, with peak efficiency of 93%. The experimentally measured efficiency has a peak value of 75% with nearly the same bandwidth compared with theory. For TM polarization (Figs. 5d, f), the peak diffraction efficiency reaches 87%, with a smaller bandwidth from 470 nm to 600 nm. For both polarizations, the measured efficiency spectra for different diffraction orders are consistent with the calculated results. Figs. 5g and h show the diffraction efficiency ($T_{-1}$) evolution with incident angles. The measured efficiencies exhibit flat-top peaks against incident angles and, at middle wavelengths around 500 nm, the angular bandwidth is the largest, covering the incident angle range from 20° to 80°. Note that, the incident angle $\theta_0$ is taken in the air region below the substrate in experiments, while in calculation the incident light is directly from the substrate region. According to Snell's law, the grazing incident angle $\theta_0$=90° in air corresponds to $\theta_0'$=41.8° in the substrate. Therefore, calculated diffraction efficiencies do not decline at the air angle $\theta_0$=90° (Figs. 5a, b), while the measured diffraction efficiencies decline when $\theta_0$>70° because at such large incidence angle, the light spot size is far larger than the sample size and the coupled light from air to substrate is dramatically reduced.

In summary, we have put forward a perfect diffraction all-dielectric metagrating formed by binary rectangular structures and demonstrated analog wavefront shaping with robust high efficiency over the full visible spectrum. Unitary diffraction efficiency can be achieved in transmission mode, with complete suppression of all specular and higher diffraction orders. The simple inclusion shape of the metagrating enables arbitrary wavefront shaping through a displacement-varying detour phase scheme for advanced applications beyond beam deflection. The robust high diffraction efficiency meta-hologram with continuously modulated phase profile was experimentally realized, demonstrating large angle/wavelength/polarization tolerance. Our findings open new routes to high-efficiency wavefront shaping with analog modulation and relaxed fabrication demands, hopefully making mass production of high-efficiency meta-holograms for daily life usage feasible.

**Methods**

**Theoretical analysis and numerical Simulations.** The theoretical analysis of the perfect diffraction metagrating with fixed refractive index was performed by RCWA and multipole expansion method. The numerical simulation of diffraction efficiencies of multiple diffraction orders of the realistic $SiN_x$ metagratings with experimentally obtained material parameters was conducted by FEM method.

**Experimental samples and measurement**.

The metagrating was fabricated on a silicon dioxide substrate. First, a layer of 300-nm-thick $SiN_x$ film was deposited on the substrate by Inductively coupled plasma-chemical vapor deposition (ICP-CVD) system (Plasmalab System 100 ICP180, Oxford). Next, the metagrating pattern was generated in the high-resolution positive resist (ZEP520A) by an electron-beam lithography (EBL) system (Vistec EBPG5000 ES) at 100 KV. Next, the reactive-ion-etch (RIE) (Oxford Instrument Plasmalab System 100 RIE180) with a mixture of $CHF_3$ and $O_2$ gases was used to transfer the pattern down to the $SiN_x$ layer. The whole process for the metagrating fabrication is shown in Supplementary Fig. S4 in detail. In the experimental measurement, a supercontinuum laser source was employed to illuminate the metagrating with varying incident angles. The diffracted holographic image was shown on a white screen, and captured by a Nikon Camera (Fig. 3a). The diffraction efficiencies for varying incident angles and varying wavelengths were obtained by collecting the power of those holographic images into a power meter (Supplementary Fig. S5).

**Acknowledgements**

This work was supported by National Key R&D Program of China (YS2018YFB110012), Natural Science Foundation of Guangdong Province (Grant 2020A1515010615), National Natural Science Foundation of China (NSFC) (Grant 11734012, 11604217, 61522504, 61420106014, 11761161002, 61775243, 51601119 and 11574218), and the Guangdong Provincial Innovation and Entrepreneurship

Project (Grant 2016ZT06D081). Z.-L.D. is also funded by the China Scholarship Council (201906785011).

**Author contributions**

Z.-L.D. initiated the idea and made the theoretical analysis. X.Y. and Z.-L.D. carried out the simulations. X.Y., Q.-A.T., H.-Y.Q., and Z.-P.Z. conducted the experiment measurement with the help of Z.-L.D and S.T.. The sample was fabricated by H.-Y.Q., Z.-P. Z., and J.-W.D.. Z.-L.D., X.Y., J.-W.D., and X.L analyzed the data. Z.-L.D. wrote the manuscript with input from X.Y., J.-W.D., A.A., and X.L. All authors contributed to discussions about the manuscript.

**Conflict of interests**

The authors declare no conflicts of interest.

**References**

[1] P. Genevet, F. Capasso, F. Aieta, M. Khorasaninejad, R. Devlin, *Optica* **2017**, 4, 139.

[2] M. Khorasaninejad, W. T. Chen, R. C. Devlin, J. Oh, A. Y. Zhu, F. Capasso, *Science* **2016**, 352, 1190.

[3] M. Khorasaninejad, A. Y. Zhu, C. Roques-Carmes, W. T. Chen, J. Oh, I. Mishra, R. C. Devlin, F. Capasso, *Nano Lett.* **2016**, 16, 7229.

[4] E. Arbabi, A. Arbabi, S. M. Kamali, Y. Horie, A. Faraon, *Optica* **2016**, 3, 628.

[5] A. Arbabi, Y. Horie, M. Bagheri, A. Faraon, *Nat. Nanotechnol.* **2015**, 10, 937.

[6] J. P. Balthasar Mueller, N. A. Rubin, R. C. Devlin, B. Groever, F. Capasso, *Phys. Rev. Lett.* **2017**, 118, 113901.

[7] Z.-L. Deng, S. Zhang, G. P. Wang, *Opt. Express* **2016**, 24, 23118.

[8] W. T. Chen, A. Y. Zhu, V. Sanjeev, M. Khorasaninejad, Z. Shi, E. Lee, F. Capasso, *Nat. Nanotechnol.* **2018**, 13, 220.

[9] S. Wang, P. C. Wu, V.-C. Su, Y.-C. Lai, M.-K. Chen, H. Y. Kuo, B. H. Chen, Y. H. Chen, T.-T. Huang, J.-H. Wang, R.-M. Lin, C.-H. Kuan, T. Li, Z. Wang, S. Zhu, D. P. Tsai, *Nat. Nanotechnol.* **2018**, 13, 227.

[10] G. Zheng, H. Mühlenbernd, M. Kenney, G. Li, T. Zentgraf, S. Zhang, *Nat. Nanotechnol.* **2015**, 10, 308.

[11] L. Wang, S. Kruk, H. Tang, T. Li, I. Kravchenko, D. N. Neshev, Y. S. Kivshar,

*Optica* **2016**, 3, 1504.
[12] Z.-L. Deng, M. Jin, X. Ye, S. Wang, T. Shi, J. Deng, N. Mao, Y. Cao, B.-O. Guan, A. Alù, G. Li, X. Li, *Adv. Fun. Mater.* **2020**, 1910610. https://doi.org/10.1002/adfm.201910610.
[13] A. Nemati, Q. Wang, M. Hong, J. Teng, *Opto-electronic Adv.* **2018**, 1, 180009.
[14] Y. Zhang, L. Shi, D. Hu, S. Chen, S. Xie, Y. Lu, Y. Cao, Z. Zhu, L. Jin, B.-O. Guan, S. Rogge, X. Li, *Nanoscale Horizons* **2019**, 4, 601.
[15] D. Wen, J. J. Cadusch, J. Meng, K. B. Crozier, *Adv. Fun. Mater.* **2019**, 1906415.
[16] Y. Ra'di, D. L. Sounas, A. Alù, *Phys. Rev. Lett.* **2017**, 119, 067404.
[17] Z.-L. Deng, S. Zhang, G. P. Wang, *Nanoscale* **2016**, 8, 1588.
[18] Z.-L. Deng, J. Deng, X. Zhuang, S. Wang, T. Shi, G. P. Wang, Y. Wang, J. Xu, Y. Cao, X. Wang, X. Cheng, G. Li, X. Li, *Light Sci. & Appl.* **2018**, 7, 78.
[19] N. M. Estakhri, V. Neder, M. W. Knight, A. Polman, A. Alù, *ACS Photon.* **2017**, 4, 228.
[20] Y. Ra'di, A. Alù, *ACS Photon.* **2018**, 5, 1779.
[21] A. M. H. Wong, G. V. Eleftheriades, *Phys. Rev. X* **2018**, 8, 011036.
[22] Z.-L. Deng, Y. Cao, X. Li, G. P. Wang, *Photon. Res.* **2018**, 6, 443.
[23] H. Chalabi, Y. Ra'di, D. L. Sounas, A. Alù, *Phys. Rev. B* **2017**, 96, 075432.
[24] A. Epstein, O. Rabinovich, *Phys. Rev. Appl.* **2017**, 8, 054037.
[25] V. Popov, F. Boust, S. N. Burokur, *Phys. Rev. Appl.* **2018**, 10, 011002.
[26] V. Popov, M. Yakovleva, F. Boust, J.-L. Pelouard, F. Pardo, S. N. Burokur, *Phys. Rev. Appl.* **2019**, 11, 044054.
[27] Z.-L. Deng, J. Deng, X. Zhuang, S. Wang, K. Li, Y. Wang, Y. Chi, X. Ye, J. Xu, G. P. Wang, R. Zhao, X. Wang, Y. Cao, X. Cheng, G. Li, X. Li, *Nano Lett.* **2018**, 18, 2885.
[28] S. Wang, F. Li, J. Deng, X. Ye, Z.-L. Deng, Y. Cao, B.-O. Guan, G. Li, X. Li, *J. Phys. D: Appl. Phys.* **2019**, 52, 324002.
[29] M. Li, L. Jing, X. Lin, S. Xu, L. Shen, B. Zheng, Z. Wang, H. Chen, *Adv. Opt. Mater.* **2019**, 7, 1900151.
[30] M. Khorasaninejad, F. Capasso, *Nano Lett.* **2015**, 15, 6709.
[31] E. Khaidarov, H. Hao, R. Paniagua-Domínguez, Y. F. Yu, Y. H. Fu, V. Valuckas, S. L. K. Yap, Y. T. Toh, J. S. K. Ng, A. I. Kuznetsov, *Nano Lett.* **2017**, 17, 6267.
[32] D. Sell, J. Yang, S. Doshay, R. Yang, J. A. Fan, *Nano Lett.* **2017**, 17, 3752.
[33] Z. Fan, M. R. Shcherbakov, M. Allen, J. Allen, B. Wenner, G. Shvets, *ACS Photon.* **2018**, 5, 4303.
[34] T. Shi, Y. Wang, Z.-L. Deng, X. Ye, Z. Dai, Y. Cao, B.-O. Guan, S. Xiao, X. Li, *Adv. Opt. Mater.*, 0, 1901389.
[35] D. Sell, J. Yang, E. W. Wang, T. Phan, S. Doshay, J. A. Fan, *ACS Photon.* **2018**, 5, 2402.
[36] X. Dong, J. Cheng, F. Fan, S. Chang, *Opt. Lett.* **2019**, 44, 939.
[37] J. Jiang, D. Sell, S. Hoyer, J. Hickey, J. Yang, J. A. Fan, *ACS Nano* **2019**, 13, 8872.
[38] A. Casolaro, A. Toscano, A. Alù, F. Bilotti, *IEEE Transactions on Antennas and Propagation* **2020**, 68, 1542.
[39] O. Rabinovich, A. Epstein, *IEEE Transactions on Antennas and Propagation* **2020**, 68, 1553.

[40] J. Du, Z. Lin, S. T. Chui, W. Lu, H. Li, A. Wu, Z. Sheng, J. Zi, X. Wang, S. Zou, F. Gan, *Phys. Rev. Lett.* **2011**, 106, 203903.
[41] A. Wu, H. Li, J. Du, X. Ni, Z. Ye, Y. Wang, Z. Sheng, S. Zou, F. Gan, X. Zhang, X. Wang, *Nano Lett.* **2015**, 15, 2055.
[42] W. Liu, A. E. Miroshnichenko, *ACS Photon.* **2018**, 5, 1733.
[43] X. Li, H. Ren, X. Chen, J. Liu, Q. Li, C. Li, G. Xue, J. Jia, L. Cao, A. Sahu, B. Hu, Y. Wang, G. Jin, M. Gu, *Nat. Commun.* **2015**, 6, 6984.

## Figures

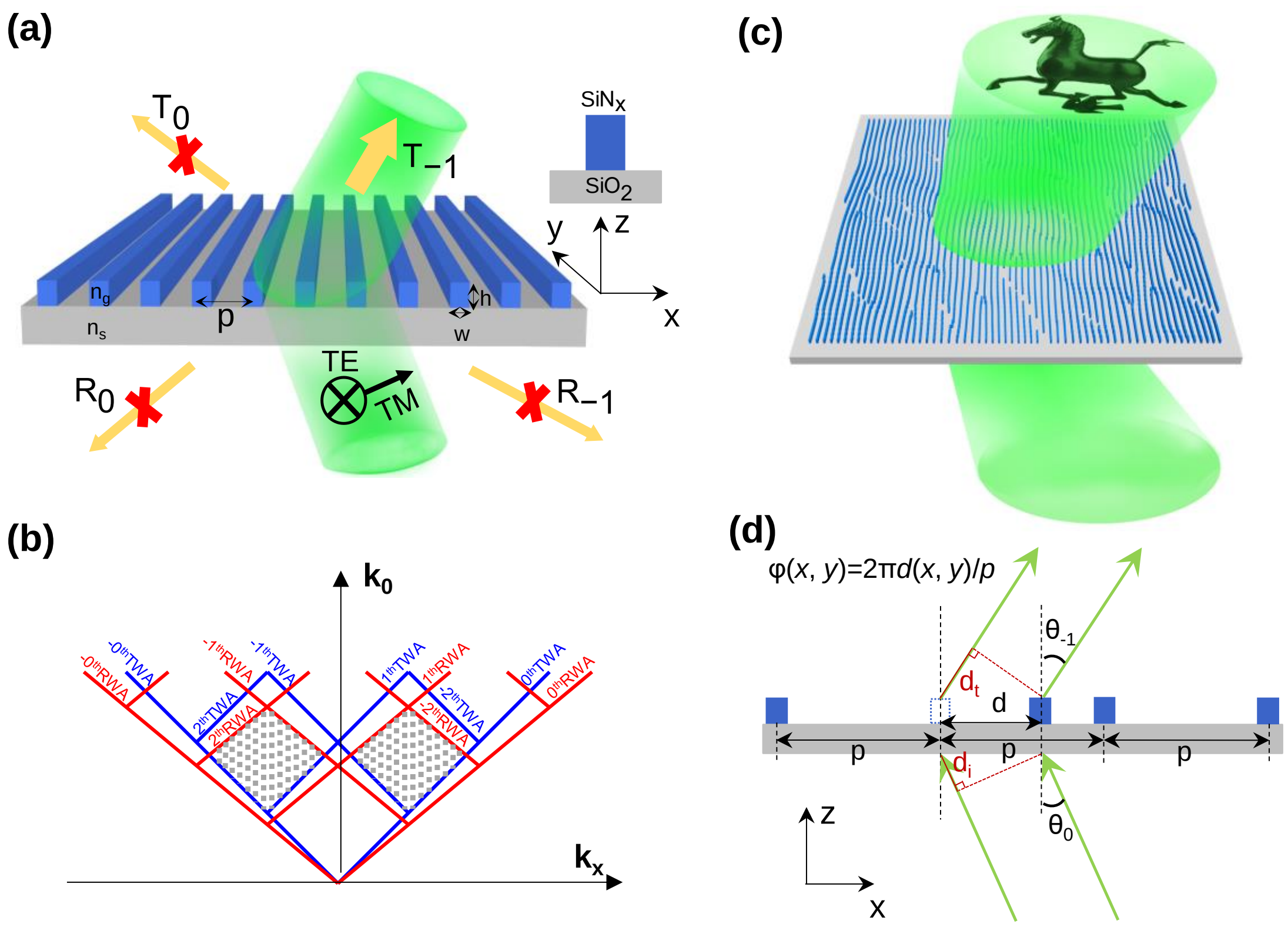

**Figure 1. Schematic of a CMOS-compatible SiN$_X$ metagrating.** (a) The metagrating is composed of SiN$_x$ nanobars with width $w$, height $h$, and periodicity $p$ on top of a SiO2 substrate, working in the diffraction regime that supports 0$^{th}$ and -1$^{st}$ diffraction channels in both reflection ($R_0$, $R_{-1}$) and transmission ($T_0$, $T_{-1}$) spaces. After proper optimization, only the $T_{-1}$ channel is allowed, with complete suppressions of $T_0$, $R_0$ and $R_{-1}$. (b) Diffraction order chart showing a series of transmissive Woods anomaly (TWA) lines (blue) and reflective Woods anomaly (RWA) lines (red). The gray dotted areas represent the working regimes for perfect diffraction. (c) Continuously modulated diffraction metagratings can perform high-efficiency wavefront shaping in the visible range. (d) Phase modulation rule for arbitrary wavefront shaping in diffraction metagratings.

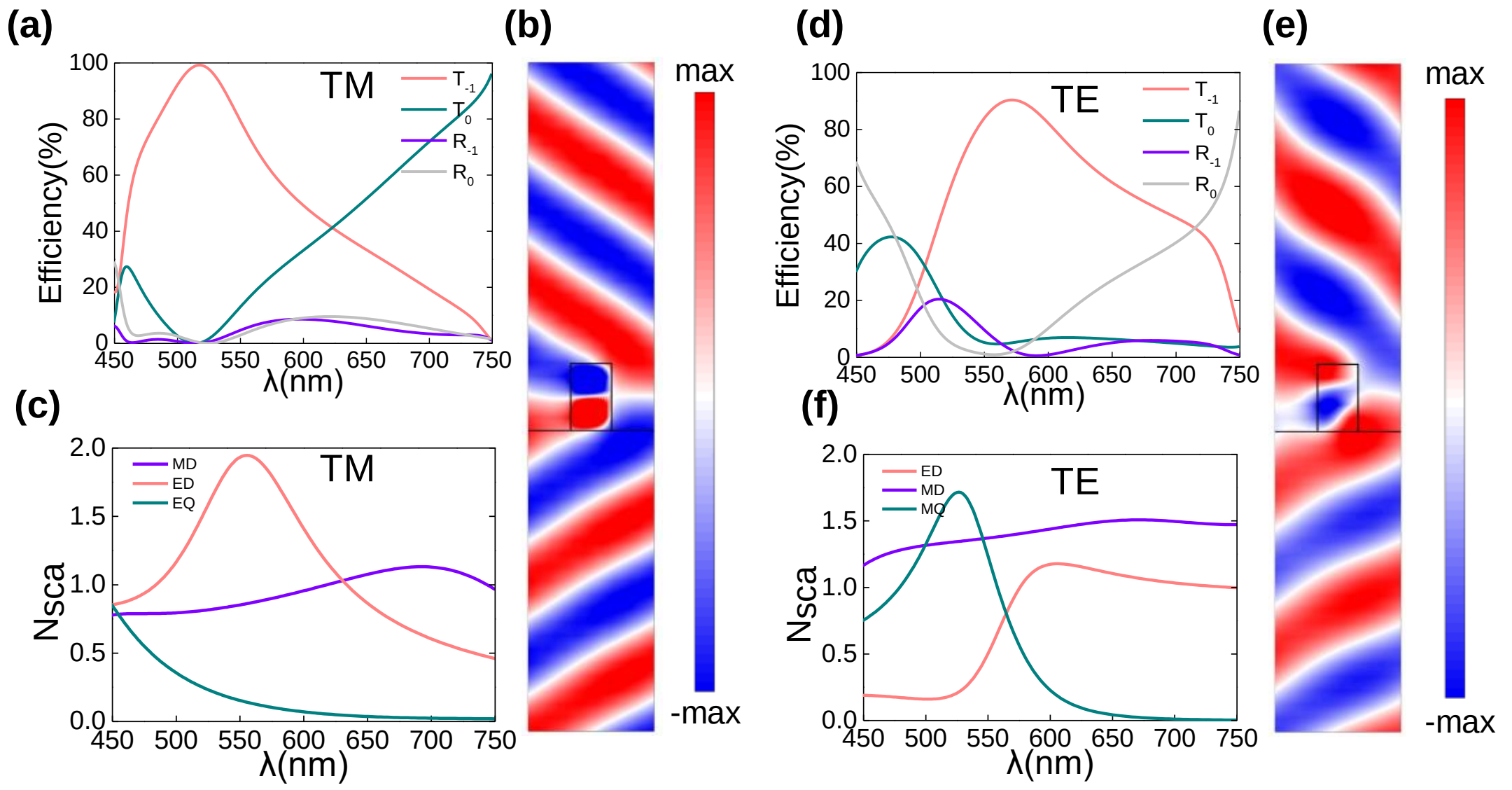

**Figure 2. Diffraction properties of the designed metagrating.** (a, d) Diffraction efficiency spectra obtained by RCWA for all diffraction channels ($T_{-1}$, $T_0$, $R_{-1}$, $R_0$) of the metagrating. (b, e) Field patterns at peak diffraction for (c) TM and (f) TE illumination. (c, f) Multipole decomposition of the $SiN_x$ nanobar scattering, dominant multipole modes include magnetic dipole (MD), electric dipole (ED), magnetic quadrapole (MQ) and electric quadrapole (EQ) for (c) TM and (f) TE polarization. The parameters of the metagrating are $h$=267.5 nm, $w$=162 nm and $p$=500 nm. The illumination angle is 30°.

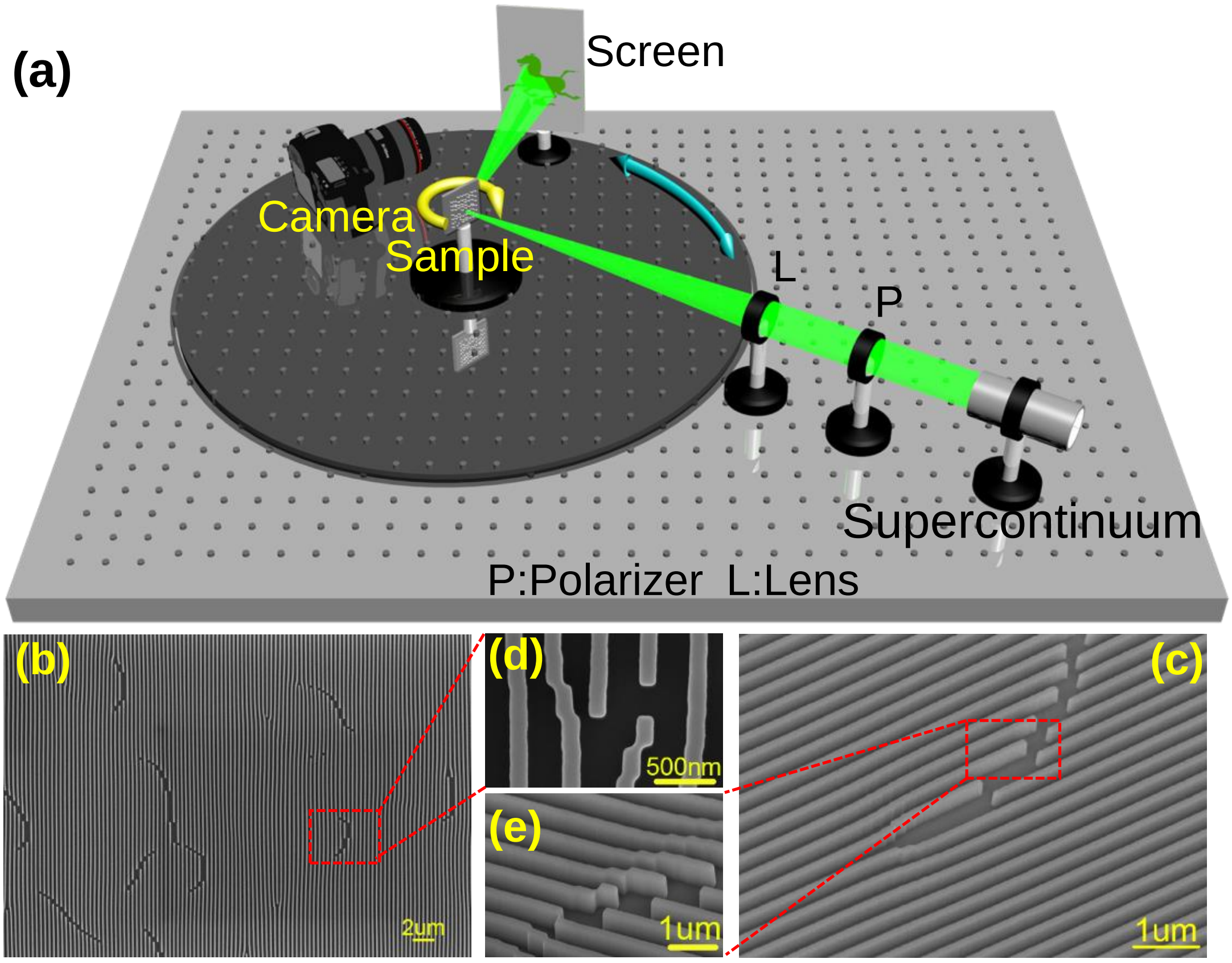

**Figure 3. Phase modulation for arbitrary wavefront shaping based on perfect diffraction metagrating.** (a) Angle-resolved optical setup for the wide-angle full visible range holographic imaging. (b-e) SEM images of the continuously modulated $SiN_x$ metagratings on top view (b, d), and oblique view (c, e).

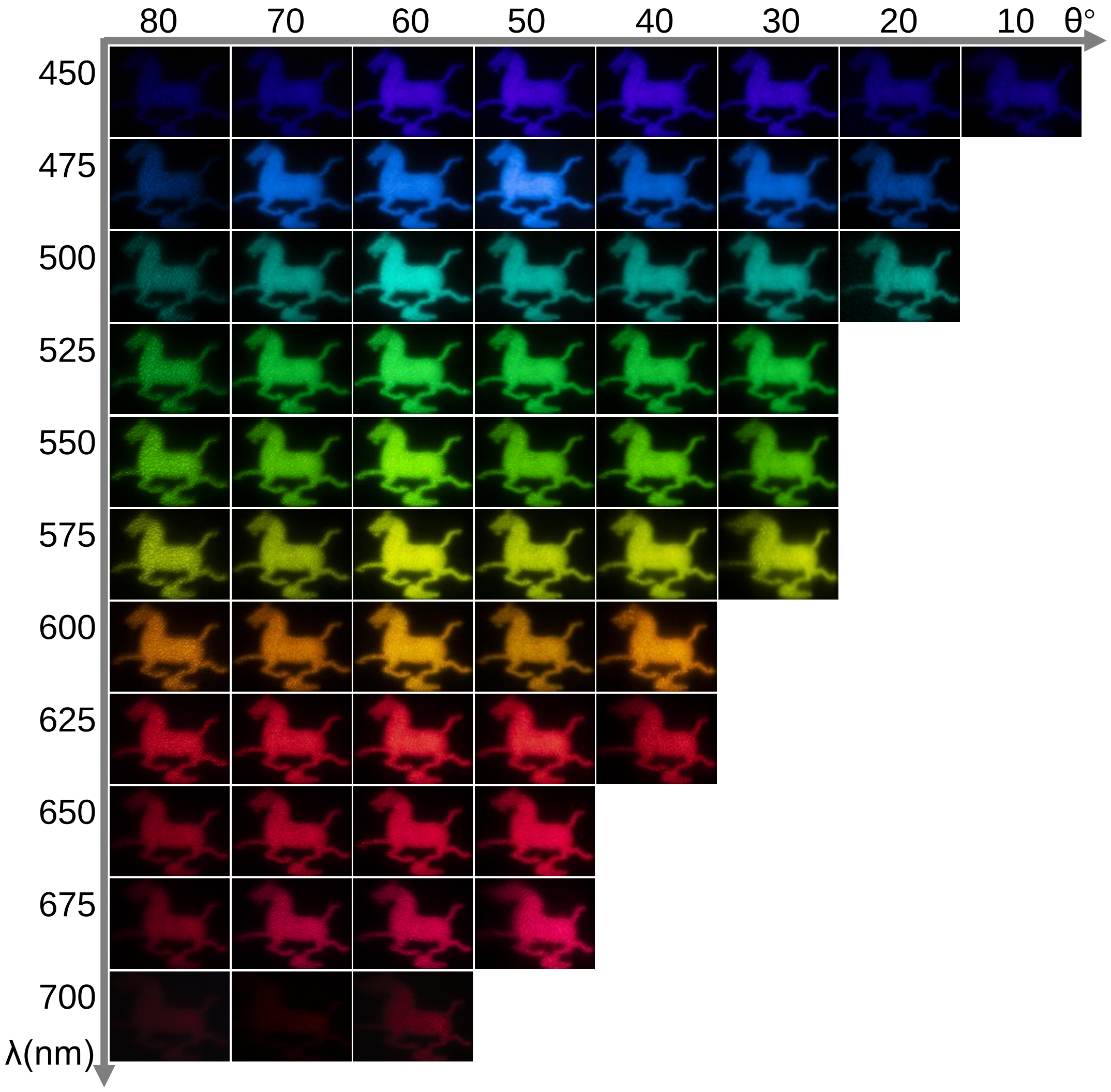

**Figure 4. Experimentally reconstructed holographic images.** Reconstructed images of the modulated metagrating are acquired at different incident angles and different wavelenths in the full visible spectrum. The illumination light is TE polarized.

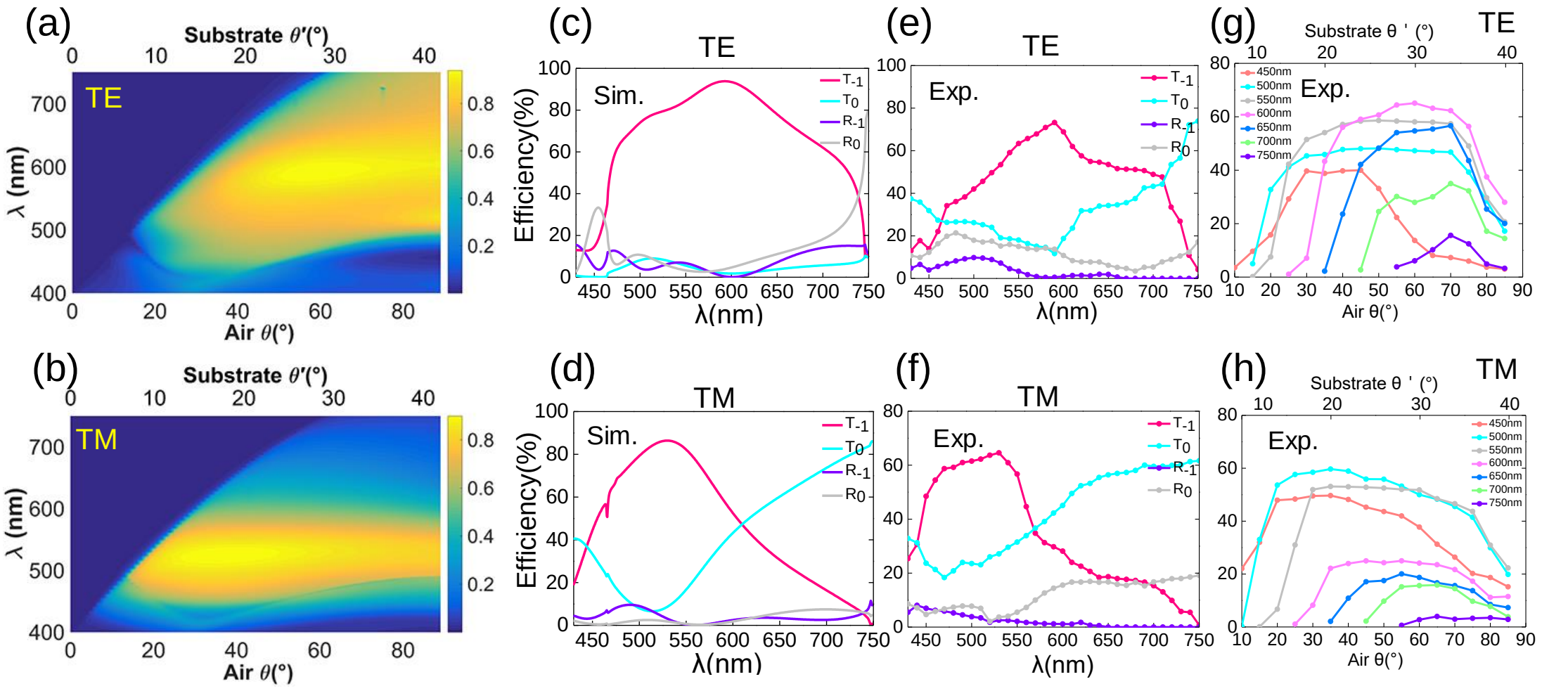

**Figure 5. Diffraction spectra of the realized metagratings** (a-b) Diffraction efficiency map for varying incident angles and wavelengths calculated with experimentally obtained material parameters, (a) for TE polarization, (b) for TM polarization, respectively. (c-d) Efficiency spectra of various diffraction orders ($T_{-1}$, $T_0$, $R_{-1}$, $R_0$) at 60° incident angle from air, obtained by calculation of realistic metagratings, for (c) TE polarization and (d) TM polarization. (e-h) Experimental measurement of diffraction efficiencies of the metagrating hologram in Fig. 3 for different wavelengths and incident angles, and for (e, g) TE polarization and TM polarization, respectively.